\documentstyle[12pt]{article}
\makeatletter
\@addtoreset{equation}{section}
\makeatother

\def\Lp{\displaystyle{\biggl(}}
\def\Rp{\displaystyle{\biggr)}}

\newcommand{\lp}{\left(}\newcommand{\rp}{\right)}

\renewcommand{\d}{\delta}

\renewcommand{\AA}{{\cal A}}

\newcommand{\GG}{{\cal G}}

\newcommand{\NN}{{\cal N}}

\newcommand{\PP}{{\cal P}}
\newcommand{\QQ}{{\cal Q}}

\newcommand{\VV}{{\cal V}}

\newcommand{\complex}{{\kern .1em {\raise .47ex
\hbox {$\scriptscriptstyle |$}}
    \kern -.4em {\rm C}}}
\newcommand{\real}{{{\rm I} \kern -.19em {\rm R}}}
\newcommand{\rational}{{\kern .1em {\raise .47ex
\hbox{$\scripscriptstyle |$}}
    \kern -.35em {\rm Q}}}
\renewcommand{\natural}{{\vrule height 1.6ex width
.05em depth 0ex \kern -.35em {\rm N}}}

\newcommand{\tr}{{\rm {Tr} \,}}

\newcommand{\pa}{\partial}
\newcommand{\pad}[2]{{\frac{\partial #1}{\partial #2}}}

\newcommand{\sla}{\raise.15ex\hbox{$/$}\kern -.57em}

\newcommand{\twiddle}{\lower.9ex\rlap{$\kern -.1em\scriptstyle\sim$}}

\renewcommand{\=}{&=&} 
\newcommand{\equ}[1]{(\ref{#1})}

\newcommand{\eq}{\begin{equation}}
\newcommand{\eqn}[1]{\label{#1}\end{equation}}
\newcommand{\eea}{\end{eqnarray}}
\newcommand{\eqa}{\begin{eqnarray}}
\newcommand{\eqan}[1]{\label{#1}\end{eqnarray}}
\newcommand{\ba}{\begin{array}}
\newcommand{\ea}{\end{array}}
\newcommand{\eqac}{\begin{equation}\begin{array}{rcl}}
\newcommand{\eqacn}[1]{\end{array}\label{#1}\end{equation}}

\renewcommand{\pad}[2]{{\displaystyle{\frac{\partial #1}{\partial #2}}}}

\newcommand{\journal}[4]{{\em #1~}#2\,(19#3)\,#4;}
\newcommand{\aihp}{\journal {Ann. Inst. Henri Poincar\'e}}

\newcommand{\ijmp}{\journal {Int. J. Mod. Phys.}}

\newcommand{\jmp}{\journal {J. Math. Phys.}}

\newcommand{\cmp}{\journal {Comm. Math. Phys.}}

\newcommand{\np}{\journal {Nucl. Phys.}}
\newcommand{\pl}{\journal {Phys. Lett.}}

\newcommand{\prep}{\journal {Phys. Reports}}

\newcommand{\annp}{\journal {Ann. Phys. (N.Y.)}}

\makeatletter
\@addtoreset{equation}{section}
\makeatother

\advance\hoffset by -0.6cm
\begin{document}
{\large    
\vspace{20mm}
\centerline{\LARGE Algebraic characterization of the Wess-Zumino} \vspace{2mm}

\centerline{\LARGE consistency conditions in gauge theory}
\vspace{9mm}

\centerline{S.P. Sorella$^1$\footnotetext[1]{Supported in part
                          by the Swiss National Science Foundation.}}
\centerline{{\small D\'epartement de Physique Th\'eorique}}
\centerline{{\small 24, quai Ernest Ansermet}}
\centerline{{\small CH -- 1211 Gen\`eve 4 (Switzerland)}}

\vspace{10mm}

\centerline{{\normalsize {\bf UGVA---DPT 1992/08--781}} }
\vspace{2cm}

\centerline{\Large{\bf Abstract}}\vspace{4mm}

\noindent
A new way of solving the descent equations corresponding to the
Wess-Zumino consistency conditions is presented.
The method relies on the introduction of an operator $\delta$ which allows
to decompose the exterior space-time derivative $d$ as a $BRS$ commutator.
The case of the Yang-Mills theories is treated in detail.
\setcounter{page}{0}
\thispagestyle{empty}

\vfill\pagebreak
\section{Introduction}
It is well known that the anomalies in gauge theories have to be non-trivial
solutions of the Wess-Zumino consistency conditions~\cite{wz}.
These conditions,
when formulated in terms of the Becchi-Rouet-Stora
transformations~\cite{brs}, yield a cohomology problem for the
nilpotent $BRS$ operator $s$:
\eq
   s \Delta = 0 \ ,
\eqn{cohm}
where $\Delta$ is the integral of a local polynomial in the fields and their
derivatives. An useful way of finding non-trivial solutions of \equ{cohm} is
given by the so-called $\it descent-equations$
technique~\cite{zumino1,zumino,dviolette,baulieu,brandt,bonora,piguet}.

Setting $\Delta = \int \AA$,${\ }$eq. \equ{cohm} translates into the local
condition
\eq
      s \AA + d\QQ = 0 ,
\eqn{cond1}
for some $\QQ$; $d$ being the exterior differential on the space-time $M$.
The operators $s$ and $d$ verify:
\eq
   s^2{\ } ={\ } d^2{\ } ={\ }sd{\ }+{\ }ds{\ }=0 \ .
\eqn{sd}
One can easily prove that equation \equ{cond1}, due to the triviality
of the cohomolgy
of $d$~\cite{dviolette,brandt,cotta}, generates a tower of descent equations
\eq\ba{lcl}
     s\QQ{\ }{\ }{\ }+{\ }d\QQ^1 = 0   \\
    s\QQ^1{\ }{\ }+{\ }d\QQ^2 = 0   \\
{\ }{\ }{\ }{\ }         ...... \\
{\ }{\ }{\ }{\ }         ...... \\
    s\QQ^{k-1}{\ }+{\ }d\QQ^k = 0   \\
    s\QQ^{k} = 0 \ ,
\ea\eqn{tower}
where the $\QQ^i$ are local polynomials in the fields.

The aim of this work is to give a systematic procedure for
generating explicit solutions of the descent equations \equ{tower}.
This will be done in the class of polynomials of differential
forms~\cite{dviolette} and for any space-time dimension and ghost-number.
The use of the space of polynomials of differential forms; i.e. polynomials
in the variables $A, dA, c, dc$ ($A$ and $c$ being respectively the gauge
connection and the ghost field), is the natural choice when dealing with
anomalies and Chern-Simons terms. This is due to the topological origin
of the latters~\cite{stora}.
It is worthwhile to mention that actually, as recently proven by
M. Dubois-Violette et al.~\cite{dvu}, the use of the space of polynomials
of forms is not a restriction on the generality of the solution of
the consistency equations.
To avoid the introduction of a reference connection~\cite{zumino} we will
assume that the principal fiber bundle for the gauge potential is trivial,
i.e. of the form ($M \times G$) with $G$ a compact semisimple Lie group.

The main idea which we will use to solve the descent equations consists
in writing the exterior derivative $d$ as a $BRS$ commutator; i.e. we will
be able to make the decomposition:
\eq
           d{\ }={\ }-[{\ }s{\ },{\ }\d{\ }]   \ ,
\eqn{dec}
where $\d$ is an operator which will be specified later. One easily shows
that, once the decomposition \equ{dec} has been found, repeated applications
of the operator $\d$ on the polynomial $\QQ^k$ which solves the last of the
equations \equ{tower} will give an explicit solution for the higher
cocycles $\QQ^{k-1},{\ }....,{\ }\QQ^1,{\ }\QQ$ and $\AA$.
One has to remember also that the form of the polynomial $\QQ^k$ is well
known~\cite{zumino,dviolette,brandt,cotta,dixon,bandelloni} and is uniquely
specified by invariant ghost monomials of the form $\tr c^k$ ( $k$ odd ).
This completes the general strategy.

We emphasize that this scheme represents an alternative way of solving the
descent equations which is completely different from the usual homotopy
setup given by the
$\it " Russian-formula "$~~\cite{zumino1,zumino,dviolette,baulieu}.
However we will see that the two schemes identify the same class of
solutions, i.e. they give expressions which differ only by a trivial cocycle.

It is interesting to note that the decomposition \equ{dec} naturally
appears in the context of the topological field theories~\cite{witten,birm}.
In this case the operator $\d$ is the generator of the topological vector
supersymmetry and allows for a complete classification of anomalies,
counterterms and non-trivial observables~\cite{silvio}.

The paper is organized as follows. In Section 2 we briefly recall some
basic properties concerning the cohomologies of $d$ and $s$. Section 3
is devoted to the study of the algebraic structure implied by the
decomposition \equ{dec}. In Section 4, after giving some explicit
examples, we present the solution of the descent equations
in the general case.

We hope that
this work will be of some help in understanding the algebraic structure which
underlies the topological nature of the anomalies.

\section{General results on the $d$ and $s$ cohomologies}

\subsection{Notations and functional idendities}
Let $\VV(A,dA,c,dc)$ be the space of polynomials of differential forms. The
$BRS$ transformations of the one-form gauge connection $A^a = A^a_\mu dx^\mu$
and of the zero-form ghost field $c^a$ are:
\eq\ba{lcl}
sA^a \= dc^a + f^{abc}c^bA^c   \ ,\\
sc^a \= {1\over 2} f^{abc}c^b c^c \ ,
\ea\eqn{brsin}
where $f^{abc}$ are the structure constants of the gauge group $G$ and $d$ is
the exterior derivative defined by
\eq
     d \omega_p = dx^\mu \pa_\mu \omega_p \ ,
\eqn{ddef}
for any $p$-form
\eq
 \omega_p = {1 \over p!}\omega_{i_1....i_p}dx^{i_1}....dx^{i_p} \ ,
\eqn{pform}
where a wedge product has to be understood.

As usual, the adopted grading is given by the sum of the form degree and of
the ghost number. The fields $A$ and $c$ are both of degree one, their ghost
number being respectively zero and one. A $p$-form with
ghost number $q$ will be denoted by $\omega_p^q$; its grading being $(p+q)$.
The two-form field strength $F^a$ reads:
\eq
 F^a = {1 \over 2} F^a_{\mu\nu} dx^\mu dx^\nu{\ }={\ }
      dA^a + {1\over 2}f^{abc}A^bA^c  \ ,
\eqn{ftens}
and
\eq
 dF^a = f^{abc}F^bA^c \ ,
\eqn{bianchi}
is its Bianchi identity.

To characterize the cohomology of $s$ and $d$ on the space of polynomials
of differential forms it is convenient to switch from
$(A, dA, c, dc)$ to another set of more natural variables.
Following~\cite{dviolette}, we choose as independent variables the set
$(A, F, c, \xi=dc)$; i.e. we replace everiwhere $dA$ with $F$ by using
\equ{ftens} and we introduce the variable $\xi=dc$ to emphasize the local
character of the descent equations \equ{tower}. Indeed, since integration by
parts is not allowed, the variable $\xi=dc$ is really an independent quantity.
On the local space $\VV(A,F,c,\xi)$ the $BRS$ operator $s$ and
the exterior derivative $d$ act as ordinary differential operators given
by
\eq
 s{\ }={\ }(\xi^a + f^{abc}c^bA^c)\pad{\ }{A^a}{\ }+{\ }
           f^{abc}{c^bc^c\over 2}\pad{\ }{c^a}{\ }+{\ }
           f^{abc}c^b\xi^c\pad{\ }{\xi^a}{\ }+{\ }
           f^{abc}c^bF^c\pad{\ }{F^a} \ ,
\eqn{soper}
\eq
 d{\ }={\ }(F^a -f^{abc}{A^bA^c \over 2})\pad{\ }{A^a}{\ }+{\ }
           \xi^a\pad{\ }{c^a}{\ }+{\ }
          f^{abc}F^bA^c\pad{\ }{F^a} \ .
\eqn{doper}
One easily checks that $s$ and $d$ are of degree one and satisfy
\eq
   s^2{\ } ={\ } d^2{\ } ={\ }sd{\ }+{\ }ds{\ }=0 \ .
\eqn{sdeq}

\subsection{The $d$ cohomology}
Even if the vanishing of the cohomology of the exterior derivative is a well
established result~\cite{dviolette,brandt,cotta} we give here a simple
proof which may be useful for the reader.

{\bf Proposition I}

The exterior derivative $d$ has vanishing cohomology on $\VV(A,F,c,\xi)$.

{\bf Proof}

The proof is easily done by introducing the filtering operator
$\NN$~\cite{dixon,bandelloni}
\eq
\NN = \xi^a\pad{\ }{\xi^a}{\ }+{\ }
      c^a\pad{\ }{c^a}{\ }+{\ }
      A^a\pad{\ }{A^a}{\ }+{\ }F^a\pad{\ }{F^a} \ ,
\eqn{filt}
according to which the exterior derivative \equ{doper} decomposes as
\eq
  d{\ }={\ }d^{(0)}{\ }+{\ }d^{(1)} \ ,
\eqn{ddec}
with
\eq
 d^{(0)}{\ }={\ }\xi^a\pad{\ }{c^a}{\ }+{\ }F^a\pad{\ }{A^a} \ ,
\eqn{d0}
and
\eq
 d^{(0)} d^{(0)}{\ }={\ }0 \ .
\eqn{nild0}
It is apparent from \equ{d0} that $d^{(0)}$ has vanishing cohomology. It then
follows that also $d$ has vanishing cohomology, due to the fact that the
cohomology of $d$ is isomorphic to a subspace of the cohomology of
$d^{(0)}$~\cite{dixon,bandelloni}.


\subsection{The $s$ cohomology and the descent equations}
The triviality of the cohomology of $d$ allows for a simple algebraic
derivation of the descent equations. To do this, let us begin by recalling
the main result on the cohomolgy of $s$.

{\bf $s$-cohomology}~\cite{dviolette,brandt,cotta,dixon,bandelloni}

The cohomology of $s$ on $\VV(A,F,c,\xi)$ is spanned by polynomials in the
variables $(c,F)$ generated by elements of the form
\eq
    \Lp \tr {c^{2m+1} \over {(2m+1)!}} \Rp{\ } \PP_{2n+2}(F) \qquad {\ }{\ }
      m,n = 1,2,....      \ ,
\eqn{scohom}
with $\PP_{2n+2}(F)$ the invariant monomial of degree $(2n+2)$; i.e.
\eq
\PP_{2n+2}(F){\ }={\ }\tr F^{n+1}  \ ,
\eqn{invpol}
where in matrix notation
\eq
  F= T^a F^a \qquad, \qquad c=T^a c^a \ ,
\eqn{matrix}
\eq
[{\ }T^a{\ },{\ }T^b{\ }]{\ }={\ }if^{abc}T^c \qquad, \qquad
 \tr T^a T^b{\ }={\ }\d^{ab}   \ ,
\eqn{lie}
$T^a$ being the generators of a finite unitary representation of $G$.

Due to the Bianchi indentity \equ{bianchi}, the invariant monomial
$\PP_{2n+2}(F)$ has also the remarkable property of being $d$-closed:
\eq
   d\PP_{2n+2}(F){\ }=0 \ .
\eqn{closed}
The triviality of the cohomology of $d$ implies then
\eq
  \PP_{2n+2}(F){\ }={\ }d\omega^0_{2n+1}  \ ,
\eqn{exact}
which, due to \equ{sdeq}, is easily seen to generate
a tower of descent equations:
\eq\ba{lcl}
     s\omega^0_{2n+1}{\ }+{\ }d\omega^1_{2n} = 0   \\
     s\omega^1_{2n}{\ }{\ }{\ }{\ }+{\ }d\omega^2_{2n-1} = 0   \\
     {\ }{\ }{\ }{\ }    ......                    \\
     {\ }{\ }{\ }{\ }    ......                    \\
     s\omega^{2n}_1{\ }{\ }{\ }{\ }+{\ }d\omega^{2n+1}_0 = 0   \\
     s\omega^{2n+1}_0 = 0   \ .
\ea\eqn{descequ}
In particular, eq. \equ{scohom} implies that the non-trivial
solution of the last equation in \equ{descequ} corresponding to
$\PP_{2n+2}(F)$ is given by the ghost
monomial of degree $(2n+1)$:
\eq
      \omega^{2n+1}_0 {\ }={\ }\tr {c^{2n+1} \over {(2n+1)!}} \ .
\eqn{omegsolut}
One has to note that the descent equations \equ{descequ} are still valid in
the more general case of an invariant polynomial $\QQ_{2n+2}(F)$ which is the
product of several elements of the basis \equ{invpol}:
\eq
   \QQ_{2n+2}(F){\ }={\ } \prod_{i=1}^{I} \PP_{m_i}(F) \ ,
     {\ }{\ }{\ }{\ }\qquad {\sum_i m_i = 2n+2}     \ .
\eqn{genpol}
In this case the corresponding non-trivial solution for the ghost cocycle
$\omega_0^{2n+1}$ is given by the more complex expression
\eq
   \omega_0^{2n+1}{\ }={\ }\prod_{j=1}^{J} \lp \tr {c^{p_j} \over {p_j}!} \rp
      \ , {\ }{\ }{\ }{\ }     \qquad {\sum_j p_j = 2n+1}     \ .
\eqn{genghost}
However, as one can easily understand, the knowledge of the solution
of the equations \equ{descequ} for the basic monomials $\PP_{2n+2}(F)$ allows
to characterize also the more general case of eq. \equ{genpol}.
We can assume then, without loss of generality,  that the descent equations
\equ{descequ} refer always to the monomials of the basis \equ{invpol}

\section{Decomposition of the exterior derivative}
The purpose of this section is to analyse in details the algebraic relations
implied by a decomposition of the exterior derivative $d$ as the one
proposed in \equ{dec}.

\subsection{ Algebraic relations}
Let us introduce the operators $\delta$ and $\GG$ defined by
\eq
\delta{\ }={\ }-A^a\pad{\ }{c^a}{\ }+{\ }
        ( F^a + f^{abc}{A^b A^c\over 2} )\pad{\ }{\xi^a} \ ,
\eqn{deltaoper}
and
\eq
\GG{\ }={\ }-F^a\pad{\ }{c^a}{\ }+{\ }f^{abc}F^bA^c\pad{\ }{\xi^a} \ .
\eqn{ggoper}
It is easily verified that $\delta$ and $\GG$ are respectively of degree zero
and one and that the following algebraic relations hold:
\eq
           d{\ }={\ }-[{\ }s{\ },{\ }\d{\ }]   \ ,
\eqn{alg1}
\eq
           [{\ }d{\ },{\ }\d{\ }]{\ }={\ }2\GG   \ ,
\eqn{alg2}
\eq
   \{{\ }d{\ },{\ }\GG{\ }\}{\ }={\ }0{\ }\qquad \ , \qquad
       \GG \GG{\ }={\ }0  \ ,
\eqn{alg3}
\eq
   \{{\ }s{\ },{\ }\GG{\ }\}{\ }={\ }0{\ }\qquad \ , \qquad
       [{\ }\GG{\ },{\ }\d{\ }]={\ }0  \ .
\eqn{alg4}
One sees from \equ{alg1} that, as announced, the operator $\d$ allows to
decompose the exterior derivative $d$ as a $BRS$ commutator. This property,
as it will be shown in the next chapters, will give a systematic procedure
for solving the descent equations \equ{descequ}. Notice that the closure of
the algebra between $d$, $s$ and $\d$ requires the introduction of the
operator $\GG$. Remarkably, this operator, already present in the work of
Brandt et al.~\cite{brandt}, here appears in a natural way.

\subsection{Properties of $\GG$ }
We will establish here some algebraic properties concerning the
operator $\GG$ which will be useful in the following.
Let us begin by showing that the action of $\GG$ on the ghost-monomial
\equ{omegsolut} gives a trivial $BRS$ cocycle.

{\bf Proposition II}

\eq
    \GG  \lp \tr {c^{2n+1} \over {(2n+1)!}} \rp{\ }={\ }s \Omega_2^{2n-1} \ ,
\eqn{ggprop}
for some two-form $\Omega_2^{2n-1}$ with ghost number $(2n-1)$.

{\bf Proof}

The proof is easily done by noticing that the general result \equ{scohom}
on the cohomology of $s$ and the relations \equ{alg4} imply that
\eq
    \GG s \lp \tr {c^{2n+1} \over {(2n+1)!}} \rp{\ }={\ }
   -s \GG \lp \tr {c^{2n+1} \over {(2n+1)!}} \rp{\ }={\ }0   \ ,
\eqn{ggproof1}
which shows that $\GG( \tr {c^{2n+1} \over {(2n+1)!}} )$ is $s$-invariant.
Moreover from \equ{scohom} it follows that
$\GG( \tr {c^{2n+1} \over {(2n+1)!}} )$ cannot belong to the cohomolgy of
$s$; hence it is trivial.

The two-form $\Omega_2^{2n-1}$ in \equ{ggprop} is easily computed by using
the expression \equ{ggoper} for the operator $\GG$ and its general form
reads
\eq
\Omega_2^{2n-1}{\ }={\ }-{ i \over (2n-3)}{1\over (2n)!} \tr F c^{2n-1} \ .
\eqn{omegaexp}
This completes the proof.

Repeating now the same argument of eq. \equ{ggproof1} one can prove that, if
$\GG \Omega_2^{2n-1}\ne 0$, one has
\eq
    \GG \Omega_2^{2n-1}{\ }{\ }+{\ }{\ }s \Omega_4^{2n-3}{\ }={\ }0 \ .
\eqn{ggseq1}
Acting again with $\GG$ on eq. \equ{ggseq1} and using properties \equ{alg3},
\equ{alg4} one gets
\eq
    s \GG \Omega_4^{2n-3}{\ }={\ }0 \ ,
\eqn{ggseq2}
from which it follows that, if $\GG \Omega_4^{2n-3}\ne 0$,
\eq
    \GG \Omega_4^{2n-3}{\ }{\ }+{\ }{\ }s \Omega_6^{2n-5}{\ }={\ }0 \ .
\eqn{ggseq3}
As one can easily understand, this process gives a tower of descent equations
between the operators $s$ and $\GG$. They read
\eq\ba{lcl}
    \GG  \lp \tr {c^{2n+1} \over {(2n+1)!}} \rp{\ }={\ }s \Omega_2^{2n-1} \\
    \GG \Omega_2^{2n-1}{\ }{\ }+{\ }{\ }s \Omega_4^{2n-3}{\ }={\ }0 \\
{\ }{\ }{\ }{\ }......  \\
{\ }{\ }{\ }{\ }......  \\
    \GG \Omega_{2n-2}^3{\ }{\ }+{\ }{\ }s \Omega_{2n}^1{\ }={\ }0 \ .
\ea\eqn{gsdescequ}
To close the tower, let us apply once more the operator $\GG$ on the last
of the equations \equ{gsdescequ}. Using \equ{alg3} and \equ{alg4} one has:
\eq
      s \GG \Omega_{2n}^1{\ }={\ }0 \ ,
\eqn{ggseq4}
from which it follows that $\GG \Omega_{2n}^1$ is a $BRS$ invariant
$(2n+2)$-form with zero ghost number.
Expressions \equ{scohom}, \equ{invpol} show then that $\GG \Omega_{2n}^1$
is nothingh but the invariant monomial of degree $(2n+2)$:
\eq
    \GG \Omega_{2n}^1{\ }={\ }(const.){\ }\PP_{2n+2}(F) \ ,
\eqn{endtower}
where the $constant$ can be computed by means of the general formula
\equ{omegaexp}. Equation \equ{endtower} closes the tower of descent equations
\equ{gsdescequ} generated by $\GG$ and $s$.

To conclude this section let us compute, for a better understanding of
equations \equ{gsdescequ} - \equ{endtower}, the $\Omega$-cocycles for the
$BRS$ invariant ghost monomials of degree three and five:
\eq
   f^{abc}{ c^a c^b c^c \over 3!} \qquad \ , \qquad
   d^{abc} f^{bmn} f^{cpq} { c^a c^m c^n c^p c^q \over 5!} \ ,
\eqn{c3c5}
where $d^{abc}$ is the invariant totally symmetric tensor of rank three
\eq
    d^{abc}{\ }={\ }{1 \over 2} \tr T^a \{{\ }T^b{\ },{\ }T^c{\ }\} \ .
\eqn{dexpr}
In the case of ${1 \over 3!}f^{abc} c^a c^b c^c $ the tower
\equ{gsdescequ} - \equ{endtower} reads
\eq
    \GG f^{abc}{ c^a c^b c^c \over 3! }{\ }={\ }s\Omega_2^1 \ ,
\eqn{c3descequ1}
\eq
    \GG \Omega_2^1 = - \PP_4(F) \ ,
\eqn{c3descequ2}
with $\Omega_2^1$ and $\PP_4(F)$ given by
\eq
     \Omega_2^1{\ }={\ }F^a c^a \qquad \ , \qquad
     \PP_4(F){\ }={\ }F^aF^a \ .
\eqn{c3solut}
For the ghost monomial of degree five the descent equations
\equ{gsdescequ} - \equ{endtower} are a little more extended
\eq
    \GG{\ }\lp d^{abc} f^{bmn} f^{cpq} { c^a c^m c^n c^p c^q \over 5! } \rp
     {\ }={\ }s\Omega_2^3 \ ,
\eqn{c5descequ1}
\eq
    \GG \Omega_2^3{\ }+{\ }s\Omega_4^1{\ }={\ }0 \ ,
\eqn{c5descequ2}
\eq
    \GG \Omega_4^1{\ }={\ }{1 \over 3} \PP_6(F) \ ,
\eqn{c5descequ3}
where $\Omega_2^3$, $\Omega_4^1$ and $\PP_6(F)$ are computed to be
\eq\ba{lcl}
     \Omega_2^3{\ }={\ }{1 \over 12}d^{abc}F^a c^b f^{cmn} c^m c^n \\
     \Omega_4^1{\ }={\ }-{1 \over 3}d^{abc} F^a F^b c^c  \\
     \PP_6(F){\ }={\ }d^{abc}F^aF^bF^c \ .
\ea\eqn{c5solut}

\section{Solution of the descent equations}
In this chapter we will apply the results established in the previous
sections to obtain in a closed form a class of solutions of the descent
equations \equ{descequ}. This will be done by using the
decomposition of the exterior derivative \equ{alg1} as well as the descent
equations \equ{gsdescequ} - \equ{endtower} involving the operators $\GG$
and $s$.

For the sake of clarity and to make contact with the solutions given by the
$"\it Russian-formula "$~\cite{zumino1,zumino,dviolette,baulieu}
let us proceed by discussing some explicit examples.

\subsection{ The case n=1}
In this case, relevant for the two-dimensional gauge anomaly and for the
three-dimensional Chern-Simons term, the descent equations \equ{descequ}
read:
\eq\ba{lcl}
     s\omega^0_3{\ }+{\ }d\omega^1_2 = 0   \\
     s\omega^1_2{\ }+{\ }d\omega^2_1 = 0   \\
     s\omega^2_1{\ }+{\ }d\omega^3_0 = 0   \\
     s\omega^3_0 = 0   \ ,
\ea\eqn{n1equ}
where, from eq. \equ{omegsolut}, $\omega^3_0$ is given by
\eq
   \omega^3_0{\ }={\ }f^{abc}{ c^a c^b c^c \over 3!} \ .
\eqn{c3sol}
Acting with the operator $\d$ of eq. \equ{deltaoper} on the last of the
equations \equ{n1equ} one gets
\eq
    [{\ }\d{\ },{\ }s{\ }]\omega^3_0{\ }+{\ }s\d \omega^3_0{\ }={\ }0 \ ,
\eqn{c3sol1}
which, using the decomposition \equ{alg1}, becomes
\eq
    s\d \omega^3_0{\ }+{\ }d\omega^3_0={\ }0 \ .
\eqn{c3sol2}
This equation shows that $\d \omega^3_0$ gives a solution for the cocycle
$\omega^2_1$ in eqs. \equ{n1equ}. Acting again with $\d$ on the equation
\equ{c3sol2} and using the algebraic relations \equ{alg1}, \equ{alg2}
one has
\eq
   s{\d \d \over 2}\omega^3_0{\ }-\GG\omega^3_0{\ }
   +{\ }d \d \omega^3_0{\ }=0  \ .
\eqn{c3sol3}
Moreover, from the previous results \equ{ggprop} and
\equ{c3descequ1} - \equ{c3solut}, we get
\eq\ba{lcl}
    \GG \omega^3_0 {\ }={\ }s\Omega_2^1 \\
     \Omega_2^1 = F^a c^a \ ,
\ea\eqn{c3sol4}
so that eq. \equ{c3sol3} can be rewritten as:
\eq
   s( {\d \d \over 2}\omega^3_0 - \Omega^1_2){\ }
      +{\ }d \d \omega^3_0{\ }=0  \ .
\eqn{c3sol5}
One sees that $(  {1\over 2}\d \d \omega^3_0 - \Omega^1_2)$ gives
a solution for $\omega^1_2$.
To solve completely the tower \equ{n1equ} let us apply once more the operator
$\d$ on the equation \equ{c3sol5}. After a little algebra we get:
\eq
   s( {\d \d \d \over 3!}\omega^3_0 - \d \Omega^1_2)
   {\ }+{\ }d( {\d \d \over 2}\omega^3_0 - \Omega^1_2 ){\ }=0  \ .
\eqn{c3sol6}
which shows that the cocycle $\omega_3^0$ can be identified with
$( {1 \over 3!}\d \d \d \omega^3_0 - \d \Omega^1_2)$.
It is apparent then how repeated applications of the operator $\d$ on the
zero-form cocycle $\omega_0^3$ and the use of the tower
\equ{gsdescequ} - \equ{endtower} give in a simple way a
solution of the descent equations.

Summarizing, the solution of the descent equations \equ{n1equ} is given by

\eq
    \omega^0_3 {\ }={\ }{1 \over 3!} {\d \d \d }\omega_0^3{\ }-{\ }
                          \d \Omega_2^1   \ ,
\eqn{sl1}
\eq
    \omega^1_2 {\ }={\ }{1 \over 2} {\d \d  }\omega_0^3{\ }-{\ }
                           \Omega_2^1   \ ,
\eqn{sl2}
\eq
    \omega^2_1 {\ }={\ } \d \omega_0^3  \ ,
\eqn{sl3}
where, using expressions \equ{c3sol}, \equ{c3sol4}, $\omega^0_3$,
$\omega^1_2$ and  $\omega^2_1$ read:
\eq
    \omega^2_1 {\ }={\ }-{1 \over 2}f^{abc}A^a c^b c^c{\ } ={\ }
                \xi^a c^a {\ }- s(A^a c^a) \ ,
\eqn{c3sol8}
\eq
    \omega^1_2 {\ }={\ }{1 \over 2}f^{abc}A^aA^bc^c - F^ac^a{\ }={\ }
                                    -dA^a c^a \ ,
\eqn{c3sol9}
\eq
    \omega^0_3 {\ }={\ }F^aA^a - {1 \over 6}f^{abc}A^aA^bA^c{\ }={\ }
      \tr ( FA + {i \over 3}A^3 ) \ .
\eqn{c3sol10}

One easily recognizes that the expressions \equ{c3sol8} - \equ{c3sol10}
coincide,
modulo an $s$ or a $d$-coboundary, with the solution given, for instance,
in the work of Zumino, Wu and Zee~\cite{zumino1}. In particular,
\equ{c3sol9} and \equ{c3sol10} give respectively the two-dimensional gauge
anomaly (modulo a $d$-coboundary) and the three-dimensional Chern-Simons
term.

\subsection{ The case n=2}
In this example, relevant for the gauge anomaly in four dimensions, the
tower \equ{descequ} takes the form:
\eq\ba{lcl}
     s\omega^0_5{\ }+{\ }d\omega^1_4 = 0   \\
     s\omega^1_4{\ }+{\ }d\omega^2_3 = 0   \\
     s\omega^2_3{\ }+{\ }d\omega^3_2 = 0   \\
     s\omega^3_2{\ }+{\ }d\omega^4_1 = 0   \\
     s\omega^4_1{\ }+{\ }d\omega^5_0 = 0   \\
     s\omega^5_0 = 0   \ ,
\ea\eqn{n2equ}
where, according to equation \equ{omegsolut}:
\eq
 \omega_0^5{\ }={\ }
   d^{abc} f^{bmn} f^{cpq} { c^a c^m c^n c^p c^q \over 5!} \ .
\eqn{n2sol1}
As in the previous case, a solution of eqs. \equ{n2equ} is easily obtained
by applying the operator $\d$ on the cocycle \equ{n2sol1}:
\eq
    \omega_5^0 {\ }={\ }{1 \over 5!}{\d \d \d \d \d }\omega_0^5{\ }-{\ }
                    {1 \over 3!}{\d \d \d }  \Omega_2^3 {\ }-{\ }
                     \d \Omega_4^1 \ ,
\eqn{n2sl5}
\eq
    \omega_4^1 {\ }={\ }{1 \over 4!}{\d \d \d \d }\omega_0^5{\ }-{\ }
                    {1 \over 2}{\d \d}  \Omega_2^3 {\ }-{\ }\Omega_4^1 \ ,
\eqn{n2sl4}
\eq
    \omega_3^2 {\ }={\ }{1 \over 3!}{\d \d \d }\omega_0^5{\ }-{\ }
                     \d  \Omega_2^3 \ ,
\eqn{n2sl3}
\eq
    \omega_2^3 {\ }={\ }{1 \over 2} {\d \d  }\omega_0^5{\ }-{\ }
                           \Omega_2^3   \ ,
\eqn{n2sl2}
\eq
    \omega_1^4 {\ }={\ }\d \omega_0^5 \ ,
\eqn{n2sl1}
where $\Omega_2^3$ and $\Omega_4^1$ belong to the tower
\equ{c5descequ1} - \equ{c5descequ3} and are given by
( see also eq. \equ{c5solut} ):
\eq
     \Omega_2^3{\ }={\ }{1 \over 12}d^{abc}F^a c^b f^{cmn} c^m c^n \ ,
\eqn{omega23}
\eq
     \Omega_4^1{\ }={\ }-{1 \over 3}d^{abc} F^a F^b c^c  \ .
\eqn{omega41}
In particular, $\omega_5^0$ and $\omega_4^1$ are computed to be:
\eq\ba{rl}
     \omega_5^0{\ }= &\!\! -{1 \over 3}d^{abc}F^a F^b A^c {\ }-{\ }
    {1 \over 120}d^{abc}f^{bmn}f^{cpq}A^a A^m A^n A^p A^q \\
  &\!\!  +{\ }{1 \over 12}d^{abc} F^a A^b f^{cmn} A^m A^n \\
    = &\!\!-{\ }{1 \over 3}
         \tr \lp F^2 A + {i \over 2}F A^3 -{1 \over 10}A^5 \rp \ .
\ea\eqn{om50sol}

\eq\ba{rl}
     \omega_4^1{\ }= &\!\! {\ }{1 \over 3}d^{abc}F^a F^b c^c{\ }-{\ }
      {1 \over 12}d^{abc}F^a c^b f^{cmn}A^m A^n \\
 &\!\!  -{\ }{1 \over 6}d^{abc}F^a A^b f^{cmn}A^m c^n {\ }+{\ }
    {1 \over 24}d^{abc}f^{bmn}f^{cpq}A^a A^m A^n A^p c^q \\
  = &\!\!{\ }{1 \over 3} c^a{\ } d
  \lp d^{abc} A^b dA^c{\ } + {\ }{1 \over 4}d^{abc}A^b f^{cmn} A^m A^n \rp \ .
\ea\eqn{om41sol}

and give respectively the
generalized five-dimensional Chern-Simons term and the four-dimensional
gauge anomaly. Again, expressions \equ{om50sol} and
\equ{om41sol} coincide, modulo a $d$-coboundary, with that of
ref.~\cite{zumino1}.

\subsection{ The general case }
It is strightforward now to iterate the previous construction to obtain
the solution of the descent equations \equ{descequ} in the
general case of an arbitrary $n$ ($n \ge 1$).

The solution of the ladder

\eq\ba{lcl}
     s\omega^0_{2n+1}{\ }+{\ }d\omega^1_{2n} = 0   \\
     s\omega^1_{2n}{\ }{\ }{\ }{\ }+{\ }d\omega^2_{2n-1} = 0   \\
     {\ }{\ }{\ }{\ }    ......                    \\
     {\ }{\ }{\ }{\ }    ......                    \\
     s\omega^{2n}_1{\ }{\ }{\ }{\ }+{\ }d\omega^{2n+1}_0 = 0   \\
     s\omega^{2n+1}_0 = 0   \ ,
\ea\eqn{descequ1}
is given by

\eq
      \omega^{2n+1}_0 {\ }={\ }\tr {c^{2n+1} \over {(2n+1)!}} \ ,
\eqn{omegsolut1}
\eq
 \omega_{2p}^{2n+1-2p}{\ }={\ }{\d^{2p} \over (2p)!} \omega^{2n+1}_0{\ }-{\ }
       \sum_{j=0}^{p-1} {\d^{2j} \over (2j)!}
         \Omega^{2n+1-2p+2j}_{2p-2j} \ ,
\eqn{even}

for the even space-time form sector and

\eq
     \omega_1^{2n}{\ }={\ }\d \omega^{2n+1}_0  \ ,
\eqn{omega12sol}
\eq
 \omega_{2p+1}^{2n-2p}{\ }={\ }{\d^{2p+1} \over (2p+1)!}  \omega^{2n+1}_0
   {\ }-{\ }\sum_{j=0}^{p-1} {\d^{2j+1} \over (2j+1)!}
         \Omega^{2n+1-2p+2j}_{2p-2j} \ ,
\eqn{odd}

for the odd sector and $p{\ }={\ }1, 2,.......n$.

The $\Omega$-cocycles in \equ{even}, \equ{odd} belong to the tower
\equ{gsdescequ} - \equ{endtower} and are computed by using the general formula
\equ{omegaexp}.

Equations \equ{omegsolut1} - \equ{odd} generalize the results of the
previous examples and show how the use of the operator $\d$
gives a simple way of generating explicit
solutions. It is easy to check indeed that, for $n > 2$,
the expressions \equ{omegsolut1} - \equ{odd} coincide, modulo an $s$ or
a $d$-coboundary, with the results obtained
in~\cite{zumino1,zumino,dviolette,baulieu}.

\section{ Conclusion }

We have presented a new way of solving the descent equations associated
with the Wess-Zumino consistency conditions. The main ingredient has been
the introduction of the operator $\d$ which decomposes
the exterior derivative $d$ as a $BRS$ commutator and which allows to
introduce in a natural way the operator $\GG$. This operator, already used
by Brandt et al.~\cite{brandt}, generates together with the $BRS$ operator $s$
a new tower of descent equations which
are easily disentangled using the general results on the cohomology of $s$.
Moreover the algebraic properties of $\d$ and $\GG$
allow for a complete characterization of the
solutions of Wess-Zumino consistency conditions.
These solutions turn out to coincide,
modulo trivial cocycles, with the ones already obtained by using the
homotopy $\it " Russian-formula "$~~\cite{zumino1,zumino,dviolette,baulieu}.

Applications to gravity and topological theories are under
investigation~\cite{werneck}.

\noindent{\large{\bf Acknowledgments}}: I am grateful to
O. Piguet and R. Stora for useful discussions and comments.


}    
\end{document}